\let \new=\newcommand
\new{\lsim}{\raisebox{-0.3ex}{\mbox{$\stackrel{<}{_\sim} \,$}}}
\new{\gsim}{\raisebox{-0.3ex}{\mbox{$\stackrel{>}{_\sim} \,$}}}
\new{\del}{\partial}
\new{\diff}{{\rm d}}
\new{\dt}{\rm{d}_t}
\new{\cM}{{\cal M}}
\new{\bhs}{{\texttt{\large m}}_{h8}}
\new{\bh}{{\texttt{\large m}}_{h}}
\new{\Msun}{M_\odot}

\documentclass{aa}
\usepackage{graphicx}
\begin{document}
\title {Formation of a proto-quasar from accretion flows in a halo}
\subtitle{}
\author{ A. Mangalam}
\institute{ Indian Institute of Astrophysics\\Koramangala, Bangalore 560034 \\
		    email: mangalam@iiap.ernet.in}
\date{}
\titlerunning{Formation of Proto-quasar}
\authorrunning{A. Mangalam}
\abstract{
 We present a detailed model for the formation of massive objects at
the  centers  of galaxies.   The effects of supernovae heating and the
 conditions of  gas loss are revisited. The escape time of the gas is
compared with the cooling time, which provides an additional condition
  not previously considered.  Its  consequences for the allowed mass
  range of the halo is  calculated and parameterized in terms of the
  spin parameter, $\lambda_v$, the  redshift of collapse, $z_c$, the
    fraction of baryons in stars, $f_\ast$, and the  efficiency of
  supernovae, $\nu$. It is shown that  sufficient gas is retained to
  form massive dark objects and quasars even for  moderately massive
  halos but a decline is expected at low redshifts.  Subsequently, a
gaseous disk forms with a radial extent of a $~$kpc, spun up by tidal
torques and  magnetized by supernovae fields with fields strengths of
  $10-100 \mu ~G$. In a  model of a self-similar accretion flow in an
     initially   dominant halo, it is shown that for typical halo
   parameters, about $10^8  M_\odot$  accretes  via  small magnetic
 stresses (or alternatively by self-gravity induced instability or by
alpha viscosity)  in $10^8$ years into a compact region. A model of a
 self-gravitating evolution of a compact magnetized disk ($r_0 \lsim
 100$ pc),  which is relevant when a significant fraction of the disk
 mass falls in, is  presented, and it has a rapid collapse time scale
 of a million years. The two disk solutions, one for accretion in an
   imposed halo potential and the other for self-gravitating disk,
  obtained here,   have general utility and can be adapted to other
 contexts like protostellar disks as well. Implications of this work
    for dwarf galaxy formation, and a
     residual large scale seed field are also breifly discussed.
}
\maketitle
\section{Introduction}
 There seems to be increasing evidence that supermassive black holes
   are at the  centers of galaxies. Dynamical searches indicate the
 existence of massive dark  objects (MDOs) in eight systems and their
   masses range from $10^6-10^{9.5} \Msun$  (Kormendy \& Richstone
1995). Although this study does not confirm that the  central  objects
 are supermassive black holes, it has been inferred that the central
mass is contained within  $10^5$ Schwarzchild radii.  
 On an  average,
  the black hole mass is a fraction, $10^{-2}-10^{-3}$, of the total
   mass of the  galaxy and of order $10^{-3.5}$ of the bulge mass (Wandel 1999). Recent observations show a strong correlation between the
black hole mass, $\bh$, from stellar dynamical estimates, and the velocity dispersion of the host bulges ($\bh \propto \sigma^\alpha$; where $\alpha$ is reported to be in the range 3.5--5; eg. Ferrarese \& Merritt, 2000). This has been supported by reverberation mapping studies of the broad line region (Gebhardt et al. 2000). The black hole masses in active galaxies as
 inferred from their  luminosities,  assuming reasonable efficiencies
  are also in the same range, $10^{6-10} \Msun$.  Arguments based on
time variability, relativistic jets and other circumstantial  evidence
  indicate that they are relativistically compact (Blandford \& Rees
1992).  Specific examples include a $\sim 20$ pc disk spining at $500$
km s$^{-1}$ which   implies a  $2 \times 10^9 \Msun$ black hole in M87
	       (Ford et al, 1994) and evidence of $10^7
\Msun$ mass in a region of 0.1 pc extent in the case of NGC 4258 (Miyoshi et al. 
1995). One remarkable fact is that there is a decline in the quasar
   population between $z=2$ and the present epoch. The presence of
quasars at high redshifts tells us that galaxy formation had proceeded
   far enough for supermassive black holes to form in the standard
 picture (Rees 1984). A detailed model of formation of these objects,
     such as the one attempted here, should address the issues of
supernovae feedback from star formation and the mechanism of efficient
  angular momentum transport in order to explain the massive active
  nuclei as early as $z=5$. In the case of MDOs, there is a need to
    explain the compact sizes of $10-100$ pc that are implied from
			  dynamical studies.

 Broadly, the two main routes to the formation of the massive central
    objects that have been proposed are through instabilities in a
   relativistic stellar cluster or gas dynamical schemes which may
involve a direct collapse of a primordial gas cloud or accretion of a
 collapsed gaseous disk. The main drawback
of the stellar cluster models is that one must assume the existence of dense 
and massive cluster at the outset; the angular momentum transport problem to arrive at this initial scenario is difficult to overcome. One gas dynamical scheme proposed by Shlosman, Begelman \& Frank (1990) involves accretion
of gas through stellar bars which have been induced either by self-gravity or
   galaxy interactions) driving the gas from 10 kpc to about a  kpc size 'disk of clouds' which further accretes by viscous dissipation due to cloud-cloud collisions. N-body
 simulations (Sellwood \& Moore 1998) have shown that the bar weakens
substantially after a few percent of the disk mass accumulates in the
 center. This mechanism may not last long enough to drive sufficient
 matter into a compact region as the bar instability is suppressed by
   the bulge at the inner Linblad resonance. Disk accretion due
to self-gravity and  magnetic fields may then be better candidates to
transport the mass to about a 100 pc size compact region. 

 Loeb \& Rasio (1994) considered the possibility that massive black holes form
directly during the intial collapse of the protogalaxies at high redshifts and
  performed smoothed particle
   hydrodynamical simulations of gas clouds. They find that inital collapse of
a protogalactic cloud leads to the formation of a rotatinally supported 
thin disk. They argue that if the viscous transport time in the disk is small
compared to the cooling time, then the disk could collapse to a supermassive star, 
else it could form a supermassive disk; they do not provide 
gas dynamical collapse model after the formation of a disk or star. They found that the gas
   fragments into small dense clumps (presumably are converted into
  stars). The important 
consequences of supernovae feedback is not considered. Eisentein \& Loeb (1995) suggest that quasars may be associated with rare systems that 
acquire low values of angular momentum from tidal torques during the 
cosmological collapse. Typically these are $10^5 \Msun$ objects and form at
redshifts $z \gsim 10$.  Since the viscous evolution times in the collapsed 
disks of these objects are comparable to star formation times, it is important
to consider the effects of supernovae feedback in detail that could disrupt 
these objects.

Natarajan (1999), considers the effect of feedback and argues that the fraction
of the gas retained is proportional to ratio of the supernovae heat input to
the binding energy of the gas in the halo. This assumption has interesting 
consequences for the Tully-Fisher relationship. The criteria for mass loss
was considered by (Mac-Low \& Ferrara 1999; Ferrara \& Tolstoy 2000)
in the context of a blow-out or a blow-away (when the mass is completely expelled) from disk galaxy with an isothermal atmosphere. The condition for the gas to be retained is that the blow-out velocity be less than the escape velocity. However, the cooling of the hot gas is not considered. In this paper,
we consider the possibility that gas cools before it escapes the halo.

 Silk \& Rees (1998) propose that $\sim 10^5 \Msun$ clouds at high redshifts undergo direct collapse in a hierarchical (bottom-up) CDM cosmology without undergoing fragmentation and star formation arguing that the conditions in primordial clouds differ from conventional molecular clouds. The angular momentum in this picture is shed by non-axisymmetric gravitational instabilities. The more massive black holes
($~10^7 \Msun$) ejects mass from the host galaxies by assumed spherical quasar winds. In the limiting case, it provides a relationship between the mass of the central hole and the host galaxy mass. The time between halo
   virialization and the birth of quasars is short compared to the
 cosmological timescale.  Haehnelt \& Rees (1993) proposed a scenario
   in which a disk forms (after turnaround and collapse) and loses
 angular momentum due to $\alpha$ viscosity in an estimated timescale
   of $10^8$ yrs.

\begin{figure}
\resizebox{\hsize}{!}{\includegraphics{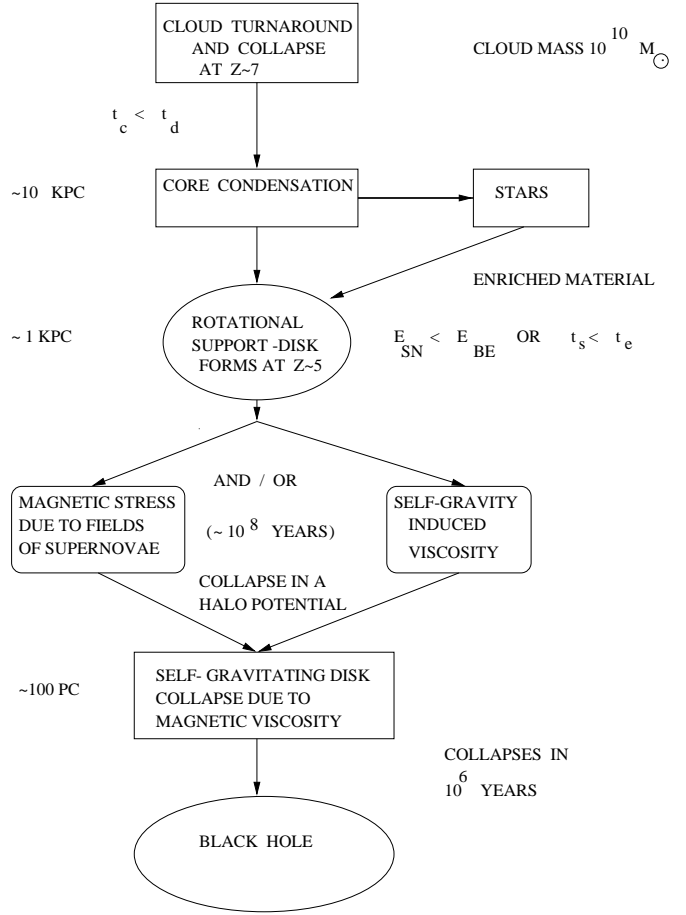}}
\caption{The evolution of a $10^{10} \Msun$ cloud leading to a formation of a
protoquasar. In first step, this cloud cools faster than the dynamical time, $t_c< t_d$. After star formation, one of the conditions is satisfied- that the supernovae heat energy is less than the binding energy, $E_{SN}<E_{BE}$ or that the cooling time of the hot gas is less than the time to escape the halo, $t_s<t_e$.} \label{picture}
\end{figure} 
 Here, we discuss a detailed physical model for the formation of protoquasars
    (or MDOs) from a magnetized accretion of a collapsed disk, the
   properties of which are obtained taking into account supernovae
 feedback in a virialized halo.  There is observational evidence that
  considerable fragmentation precedes quasar activity and the broad
 emission lines in quasar spectra indicate high metallicity (Hamann \&
 Ferland 1992).  We assume, therefore, significant star formation and
 supernovae activity occurs after the cloud, which is spun up by tidal
 torques, contracts to a radius where self-gravity is significant. The
 paper is composed of the following parts (See Fig \ref{picture})- 

\begin{enumerate}
\item 
\noindent
The formation of a gaseous disk
 with a radial extent of about a kpc, in a host galaxy as limited by supernovae feed back.  We
  investigate in \S \ref{sphere}, the range in halo mass for a given
 redshift that still retains the hot gas. The effect of the
   evolution of gas on the collisionless dark matter system is neglected.
\item
\noindent
In previous work, gravitational instabilities (examined in \S \ref{grav})
in the disk was considered as the main source of viscosity. In \S \ref{magnetic},
 justification is made for a magnetic viscosity from supernovae fields
 and the estimated accretion rate turns out to be significant; also the large scale field strength derived from the
 effective seed of small scale fields is used to explain observations (\S \ref{bdisk}).
\item
\noindent
 The collapse of the disk is calculated with a generalized viscosity
    prescription (which includes the individual cases of magnetic,
  $\alpha$ and self-gravity induced instabilities, \S \ref{viscosity}) under a halo
    dominated gravitational potential (\S \ref{bgd}) into a compact central region at rapid rate of about a $\Msun~{\rm yr}^{-1}$. A self-gravitating magnetized disk solution for this central
 object that collapses in $10^6$ yrs, is presented in \S \ref{self}. There is summary and discussion in \S \ref{disc} and conclusions in \S \ref{conclude}.

\end{enumerate} 
  
\section{The virialized spherical halo and formation of the gaseous disk}
\label{sphere}
We assume a standard spherical model for the formation of a virialized
   spherical halo to begin with, in which the gas cools to form the
  disk. Subsequently, we include in the calculations the supernovae
  heating and consider the conditions under which sufficient gas is
		      retained to form the disk.

A particularly simple and useful version of the spherical model below
assumes that the matter distribution is symmetric about a point and is
   a pressureless fluid. The shell enclosing mass of the overdense
 region, $\cM$, initially expands with the background universe, slows
      down, reaches a maximum radius before it turns around, and
 collapses. The collapse proceeds until a time when it reaches virial
equilibrium. For an average density contrast, $\delta$, and using the
fact that the background density, $ \rho_b \propto t^{-2}$ for a flat
   cosmological model, one can make the following estimates of the
typical parameters of the collapsed object (Padmanabhan \& Subramanian
		1992, Padmanabhan 1993, Peebles 1980)
\begin{eqnarray}
\rho_0&=&9.1 \times 10^{-30}~ \Omega ~h^{2}_{70} ~{\rm g~ cm}^{-3} \nonumber\\
\tau_0 &=& 0.94 \times 10^{10} h^{-1}_{70} ~{\rm yr} \nonumber \\
r_v &=& r_t/2= (78/ \delta_0) ~\cM_{10}^{1/3} ~h^{-2/3}_{70} {\rm kpc}
\\
\tau_c &=& 2.19~ \tau_0/\delta_0^{2/3} \nonumber\\
\delta_0&=& (3 \pi/2)^{3/2} ~(1+z_c)=1.686 ~(1+z_c)=1.062~(1+z_t), \nonumber
\end{eqnarray}
  where $\rho_0$ and $\tau_0$ are the current values of the comoving
density and time, the subscripts $t$ and $c$ indicate turn around and
collapse values, and $r_v$ is the radius of virialization. For a given
      model of the cosmological evolution of the initial density
 perturbation, one obtains values for $\delta_0$ or alternatively one
     can specify the collapse redshift.  As an example, $z_t=6.3$
 corresponds to $z_c=3.6$ and a $\delta_0 =$7.8.  We take $\Omega_b =
0. 1~h_{70}^{-2}$ and the mass in baryons, $M_9$, in units of $10^9
\Msun$, is related to the total mass by
\begin{equation}
   M_9 ={h_{70}^{-2}\over \Omega} ~ \cM_{10}\equiv {f_b \over 0.1}~
\cM_{10}.
\label{gas}
\end{equation}
The baryonic mass, $M_9$, will fuel a black hole, after it has formed,
   at a rate limited by the Eddington luminosity. If the accretion
  proceeds at a tenth of this rate, then the luminosity in units of
		    $10^{45} {\rm ergs s}^{-1}$ is
\begin{equation}
			  L_{45}= 1.3 ~\bhs,
\end{equation}
 where $\bhs$ is the mass of the black hole in units of $10^8~\Msun$.

\subsection{Calculation of the collapse factor $r_v/r_d$}
\label{collapse}
The gas in the massive dark halo of size $r_v$ radiative cools to form
   a disk of radial extent $r_d$. Before we consider the details of
   cooling in \S \ref{halo}, we first need to estimate the collapse
  factor, $r_v/r_d$, which is based upon the conservation of angular
      momentum. The disk forms when the gas becomes rotationally
supported. The protoquasar acquires its spin from tidal torques of its
   neighbors and N-body simulations (Barnes \& Efsthaiou 1987) and
 analytical studies (Heavens \& Peacock 1988) indicate that the spin
 parameter of the virialized system, $\lambda_v$, to be in the range
0.  01--0.1. If the angular momentum is conserved and there is no
  exchange between the gas and dark components, the ratio of angular
  momentum in the gas to the halo would remain as $M_d/\cM$, so that
\begin{equation}
   {\lambda_d \over \lambda_v}= \left ({E_d \over E} \right )^{1/2}
			     ~\left ({\cM
\over  M_d} 
\right 
			       )^{3/2}
\end{equation}
 where $\lambda \equiv L |E|^{1/2} /G M^{5/2}$ is the spin parameter,
	      $L$ is the angular mometum, and $E=-k_1 G
\cM^2/r_v$ 
 and $E_d=-k_2 G M_d^2 /r_d$ are the binding energies in the halo and
    the disk respectively. If the halo has a constant density then
$k_1=0.3$ and if it were a truncated isothermal sphere then $k_1=0.5$.
Using the form for the circular velocity, $v_c= \sqrt{G \cM/r_v}$, of
a disk spinning in an isothermal halo and taking the angular momentum
of the disk to be equal to $s M_d r_d v_c$, where $s$ is a geometrical
factor (typically of order unity; $s=2$ for an exponential disk) that
depends on the mass distribution in the disk, the radial extent of the
			   disk is given by
\begin{equation}
    r_d = r_v ~ \left ({1 \over s \sqrt{k_1}} \right) ~\lambda_v.
\end{equation}
  Typically, the disk size is a tenth of the virial radius, and the
  collapse factor ranges from 10 to 20.  For convenience we make the
 			following definitions
\begin{equation}
		    x_v= {r_v \over 10 ~{\rm kpc}}
\end{equation}
\begin{equation}
		    x_d= {r_d \over 1 ~{\rm kpc}}
\end{equation}
		where $r_d$ is the radius of the disk.
\subsection{Cooling}
\label{halo}
  In order that gaseous disks (with collapse factor estimated above)
form in the halo where star formation and supernovae take place, it is
 important to examine whether the gas can be retained in the hosts in
the first place.  In this section, we examine the constraints on black
hole hosts (those that can retain the gas) from star formation and gas
 loss due to supernovae.  Consider a virialized halo which contracts
due to cooling to a radius where it fragments due to self-gravity and
star formation takes place. Below we list the conditions that specify
 when a halo condenses to form stars, whether the gas becomes unbound
due to supernovae heating and finally whether the gas cools before it
can escape and hence remains trapped in the halo. The first two of the
 conditions were considered earlier and in this paper we introduce a
	 third necessary condition previously not considered.

\noindent {\bf C0.} Following earlier work (eg. 
 White \& Rees 1979, Rees \& Ostriker 1977, Silk 1977), we state the
	  condition that a luminous core can form in a halo
\begin{equation}
			      t_c < t_d
\label{cond}
\end{equation}
 where $t_c$ is the cooling time and $t_d$ is the dynamical time. The
heating process by was examined in some detail by Dekel \& Silk (1986,
 DS) in the context of dwarf galaxies; they found that a condition of
    gas loss amounted to the virial velocity being below a certain
 critical velocity.  At a time, $t_f$, when the hot gas in supernovae
  shells significantly fills up the volume under consideration, the
	      following constraints should be satisfied-

\noindent {\bf C1.} As given by DS, the effective heat input 
			   by supernovae is
\begin{equation}
	      E_{sn} (t_f) \ge \frac{1}{2} f_g \cM v_c^2
\label{virial}
\end{equation}
  where $f_g$ is the gas fraction in the total mass and $v_c$ is the
   circular (or virial) velocity in the halo. This implies that the
supernovae heat input should be greater the binding energy for the gas
			      to escape.

\noindent {\bf C2.} We find that an additional condition 
 for gas loss is necessary, namely, that the time for the hot gas to
   cool should be longer than the escape time, $t_e$ of the system
\begin{equation}
		    {E_{sn} \over \dot{E}} > t_e,
\end{equation}
	    where $t_e$ is approximated by the time scale
\begin{equation}
   t_e = \int_0^{r_v} {\diff r \over \sqrt{2(\overline{E}-\Phi)} },
\label{tesc}
\end{equation}
where $\overline{E}$ represents the mean kinetic energy per unit mass
  and $\Phi$ is the mean gravitational potential per unit mass.  In
 other words, even if supernovae heat input causes the gas to become
 unbound, it can still be trapped in the halo if it cools faster than
 the time required to escape, which is of the order of the dynamical
				time.
 
In order to quantify these physical constraints, we have to calculate
    $E_{sn}$, the effective heat input by supernovae. The standard
 evolution of a supernova remnant goes through two phases- adiabatic
 and radiative. In the adiabatic phase (Sedov-Taylor), the radiation
  loss is negligible and the time at which the shock front radiates
     about three-quarters of its initial energy is given by DS as
\begin{equation}
	 t_{rad} \approx 1.4 \times 10^5 ~n^{-1/2}~{\rm yr},
\label{trad}
\end{equation}
 where $n$ is the ambient hydrogen number density in cm$^{-3}$. As a
  result, the input into the gas equals the initial energy minus the
  radiative losses in the adiabatic phase and subsequently for later
  times, the gas is cooled by expansion. The cumulative energy input
 from the supernovae at a given time, $t$, will be dominated by those
   that have exploded within a time $t_{rad}$ before $t$. The star
formation rate is taken to be a constant and is approximately, $f_\ast
  M/t_d$, where $f_\ast$ is the baryonic mass fraction in stars and
   further, it is assumed that the star formation abruptly ends at
$t=t_d$. The IMF of the solar neighborhood gives rise to one supernova
 per 200 $\Msun$ of newly formed stars.  If $\nu_{200}$ is the number
       of supernova explosions per $200 \Msun$ of newly formed
stars\footnote[1]{An analytic fit to the IMF (eqn(1.3.10), Shapiro \&
 Teukolsky, 1983) which is based on Bahcall \& Soniera (1980), yields
the ratio of the number of objects in the mass range $5 < M/\Msun<30$
 to the total mass in all the stars (with a cutoff at 0.1 $\Msun$) to
   be $5 \times 10^{-3}$ per solar mass; this we have adopted as a
  fiducial value of the number of supernovae per solar mass of newly
formed stars.} with each explosion releasing $\epsilon_0=10^{51}$ ergs
of initial energy, then the energy input into the gas is then given by
\begin{equation}
    E_{sn}(t)= \epsilon_0 ~f_\ast ~{M \over 200 \Msun}~ \nu_{200}
		      ~{t_{rad} \over t_d} ~f(t)
\label{sn}
\end{equation}
and the dynamical time for the system is a quarter of the oscillation
			   period given by
\begin{equation}
  t_d = {\pi \over 2 }~ \sqrt{ r_v^3 \over G \cM} = 4.5 \times 10^8
		~\cM_{10}^{-1/2} ~x_v^{3/2}~{\rm yrs},
\label{td}
\end{equation}
   which is assumed to set the timescale for star formation. Also,
 $f(t)$, is a function of time that is of order unity and its form is
		       given by eqn (45) in DS
\begin{equation}
	 f(t)= \left \{ \begin{array}{cc} (t/t_{rad}) [1-0.14
		       (t/t_{rad})^{17/5}] & t
\leq t_{rad} \\
0.86 +0.58[(t/t_{rad})^{0.38} -1] & t > t_{rad} \end{array} \right.
\label{f(t)}
\end{equation}
  The ratio $t_{rad}/ t_d \approx 0.01~\sqrt{f_b/0.1}$ implies that
$f(t)$ is of order unity and $0 < f(t) < 3.6$ (for $0 < t < t_d$). The
   heat input from the supernovae that have exploded within a time,
 $t_{rad}$, of a given instant are the most effective in contributing
 to the heating.  Now the shells of the supernovae will start filling
up the volume. The time at which hot gas has a filling factor of order
	unity ($t=t_f$), is estimated in the following manner.

\subsection{Filling factor of SNR}
\label{shell}
 We use the well-known simple expressions (eg. Spitzer 1978) for the
advance of the supernova through the Sedov-Taylor and snow-plow phases
in a medium whose number density of hydrogen is typically $n=(1/130) (\cM_{10}/x_v^3)=(1/130)
  ({4.6 \over 1+z_c})^3 $, which is in the range of 0.01 to 1 H atom
cm$^{-3}$. In the Sedov-Taylor phase, during which the total energy in
 shock front is conserved, the advance of the shock front is given by
\begin{equation}
	    r_{st} = 10^{15} ~t^{2/5} ~n^{-1/5} ~{\rm cm},
\end{equation}
where $t$ is seconds. This phase ends when the temperature falls below
  $10^6 K$ and radiative losses are significant at a time $t_{rad}$,
 given by eqn (\ref{trad}). The radius at beginning of the radiative
	       phase as given by this condition is then
\begin{equation}
		   r_{rad}= n^{-2/5} ~40 ~{\rm pc}
\label{size}
\end{equation} 
 Next, the radiative (snow-plow) phase follows, in which the momentum
 is roughly conserved and the shock front advances according to $r_s
\propto t^{0.31}$ (Chevalier 1974).  The star formation is expected to
 occur when the gas in the halo shrinks to some size $R$ ( $\sim f_b
 r_v$ if it is isothermal or $\sim f_b^{1/3} r_v$ if the gas cloud is
 uniform) and fragments due to self-gravity (Larson 1974). Taking the
 typical size of the remnant as given by eqn (\ref{size}), the total
    shell volume of the supernovae remnants in units of the volume
			occupied by the gas is
\begin{eqnarray}
F(t)& =& {t \over t_d}~{ r_{rad}^3 ~\nu_{200}~f_\ast (M/200 \Msun) \over
		      R^3} \nonumber \\&\simeq& 10 ~{4 f_\ast
\over 1-f_\ast} ~{\nu_{200}}~n^{-1/5}~{t \over t_d},
\end{eqnarray}
where eqn(\ref{size}) and the density of the gas, $M (1-f_\ast)/(4 \pi
  R^3/3) =1.4 ~n~m_p$ was used. Since the star formation shuts off at
 $t=t_d$, the maximum value of the ratio of the total shell volume to
	       gas volume is $F(t>t_d)=10~ (4 f_\ast /
   (1-f_\ast))~{\nu_{200}}~n^{-1/5}$. This implies that the filling
factor of the supernovae shells in the galaxy will be weakly dependent
 on number density. One can take into account the porosity (which can
be thought of as the complement of the probability that a given point
in the volume is outside the $N= \nu_{200}~f_\ast ~M_9~ (5\times 10^6)
$ remnants of fractional volume, $q=F(t)/N$, occupied by one shell) by
    $Q= 1-(1-q)^N$, where $Q$ is the filling factor, which is well
    approximated in the Poisson limit of large $N$ by the formula,
\begin{equation}
			Q(t)=1- \exp{(-F(t))}.
\end{equation}
    It is clear that $Q$ at large times is close to 1 (and nearly
independent of density) and the hot gas fills the volume. In order to
 estimate the total energy input into the medium in eqn(\ref{sn}), we
calculate the time, $t_f$, when $Q \approx 1, F(t_f)=3$ which leads to
	$t_f = F(t_f) t_d/F(t_d) \simeq 0.3 t_d=30 ~t_{rad}$.

\subsection{Gas loss criteria}
\label{loss}
		    The solution leads to $f(t_f)
\approx 3$, from eqn(\ref{f(t)}). So the energy input into gas from supernovae works out to be
\begin{equation}
    E_{sn}(t_f)= 3 \times 10^{55} ~ {f_\ast \over 0.2} ~{f_b \over
0.		  1}~\nu_{200}~\cM_{10}~{\rm ergs}.
\end{equation}
		    The escape energy is given by
\begin{eqnarray}
  {1 \over 2}~f_g~\cM ~ v_c^2 &=& {1 \over 2}~{G f_g \cM^2 \over r_v} 
\nonumber \\&=&4
\times 10^{55} ~ {(1-f_\ast) \over 0.8} ~{f_b \over 0.1}~{\cM_{10}^2
\over x_v}~{\rm ergs}
\end{eqnarray}
				where
\begin{equation}
       v_c=0.7 \times 10^7 \sqrt{\cM_{10}/x_v} {\rm cm~s}^{-1}
\label{esc}
\end{equation}
 is the virial velocity. As a result the condition C1 for gas removal
			  can be written as
\begin{equation}
    {v_{sn}^2 \over v_c^2} = {3 f_\ast \over 1-f_\ast} ~{x_v \over
\cM_{10}} ~\nu_{200} >1
\label{c1}
\end{equation}
 where $v_{sn}^2= 2 E_{sn}/ (f_g \cM)$ where $f_g =1- f_b (1-f_\ast)$
 is the gas fraction in the halo mass not converted to stars.  It is
clear that enhancing the star fraction increases the energy input into
    a smaller gas fraction and hence $v_{sn}^2$ is a monotonically
  increasing function of $f_\ast$.  Also, halos of the same mass had
	deeper potential wells in the past which trap the gas
 better. Similarly, more massive halos clearly have deeper wells and
	 gas loss is less likely as seen from eqn(\ref{c1}).

 In appendix \ref{core}, we calculate the difference in gravitational
 binding energies in the initial configuration of an isothermal halo
 and a final one consisting of an exponential gaseous disk in a halo
consisting of stars and dark matter.  The gas taken to be roughly near
		       the virial temperature,
\begin{equation}
T_v= {G~ \cM ~\mu~ m_p \over 3~ k_b~ r_v}= {\cM_{10} \over x_v}~ 10^5
				  ~K
\end{equation}
   where the mean molecular weight, $\mu=0.6$. The cooling proceeds
 through line and free-free emission and we take a cooling function,
       $\Lambda(T) = 10^{-23}~{\rm ergs~s}^{-1} ~{\rm cm}^{-3}
\Lambda_{23}(T)$, provided by Sutherland \& Dopita (1993) for a
 metallicity of [Fe/H] $=-4$. The cooling rate in the virialized halo
	      during the contraction is roughly given by
\begin{equation}
\dot{E}_h = \Lambda ~n_e~N_e = 4.5 \times 10^{41} ~ \Lambda_{23}(T_v)~M_9^2 
		    ~x_v^{-3} ~{\rm ergs~ s}^{-1},
\end{equation}
   where $N_e$ is the number of electrons. Now, we can estimate the
cooling time taken for the system to cool and fragment, given that the
    source of thermal energy in the gas is one-half the change in
 gravitational potential energy from the virial theorem. The cooling
		       time, $t_c$, is given by
\begin{equation}
	      t_c = { \Delta W /2 \over \dot{E}_h(T_v)}
\end{equation}
 where $T_v = 10^5 \cM_{10}/x_v$ and plugging in the expressions for
      $\Delta W$ from appendix \ref{core}, the condition of core
	      condensation C0, thus, can be expressed as
\begin{equation}
\left (0.08 \over f_g \right) {\sqrt{ \cM_{10} x_v} \over 94}~
	  \left [{0.3 \over \lambda_v} -{1 \over 2} \right]
 		     \Lambda_{23}(T_v)^{-1} < 1.
\label{c0}
\end{equation}

\begin{figure}
\resizebox{\hsize}{!}{\includegraphics{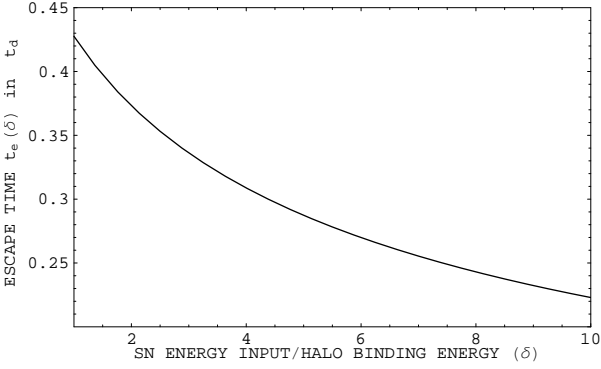}}
\caption{The dependence of the escape time of 
   the hot gas, in units of the dynamical time in the halo which is
defined as one-quarter oscillation period, on $\delta=v_{sn}^2/v_c^2$,
  the supernovae energy input in units of the halo binding energy. A
  factor of 10 in the heat input reduces the escape time, $t_e$, by
		       factor of 2.} \label{te}
\end{figure}

\begin{figure}
\resizebox{\hsize}{!}{\includegraphics{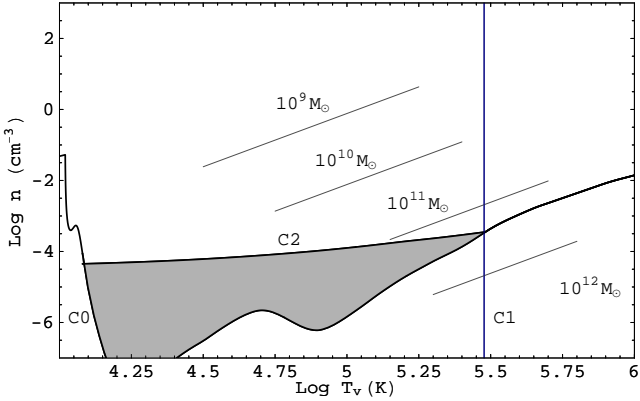}}
\caption{The shaded region bounded by the constraints, C0, C1, and 
  C2 in the cooling diagram contains the halos that have cooled and
   contracted to a radius of fragmentation; however, the supernovae
    heated gas in these systems have escape times shorter than the
 cooling time, resulting in gas-poor galaxies. The parameters chosen
 here are $f_\ast=0.2, \nu_{200}=4$, and $\lambda_v=0.075$. The halos
 above C2 for $\delta>1$ (on the left of C1) and above C0 for $\delta
  <1$ can trap the gas and hence are the candidate hosts of central
 		    massive objects.} \label{cool}
\end{figure}

 After fragmentation and star formation, the gas is heated up and the
 hot gas cools in a time, $t_s=E_{sn}/\dot{E}(T_{sn})$ where $T_{sn}=
 v^2_{sn} \mu m_p/(3 k_b)= \delta\,T_v =10^5 \,\delta \,\cM_{10}/x_v$
 is the temperature of the heated gas and $\delta =v_{sn}^2 / v_c^2=
			   (x_v \nu_{200}/
\cM_{10} )~3 f_\ast/ (1-f_\ast)$. Now, 
    we need to compare this to the escape time, as defined in eqn
  (\ref{tesc}), when the condition C1 is satisfied or a quarter of the
   oscillation period when the system is bound. Now, $\overline{E}=
E_{sn}/ (f_g M) -v_c^2/2 = v_c^2 (\delta -1)/2$, is the mean energy of
  a gas particle in the system and is zero when the escape velocity
 equals the critical velocity. Using the potential for an isothermal
  sphere that is truncated at $r_v$, $\Phi(r<r_v)= v_c^2 (\ln(r/r_v)
-1)	       $, and $r_v/v_c= (2/\pi) t_d$, we obtain
\begin{equation}
		t_e(\delta) = t_d \sqrt{2 \over \pi}
\exp({\frac{1 + \delta }{2}}) \left [ 1 - {\rm Erf}\left 
  ( \sqrt{\frac{1 + \delta}{2}} \right ) \right ] ~ \delta \geq 1
\end{equation}
 which is valid for $\delta >1$, for which the supernova heat input,
 $E_{sn}$, is such that the resulting mean energy of the particles is
  positive.  These particles can escape only if the cooling time is
  longer than the escape time. The escape time, $t_e(\delta)$, is a
      slowly decreasing function and is of order $t_d$ (See Fig
\ref{te}). For $\delta <1$, the gas remains bound and only the
condition of core condensation, C0, applies.  Hence, the condition C2,
		  $t_s/t_e >1$, can be expressed as
\begin{equation}
\left (0.08 \over f_g \right) {\sqrt{ \cM_{10} x_v} \over 94}~
 \delta ~ \Lambda_{23}(T_{sn})^{-1} {0.418~t_d \over t_e(\delta)} >1,
\label{c2}
\end{equation}
 where the effective dynamical timescale at $\delta=1$ is taken to be
  $t_d$ and $t_e(1)= 0.418 ~t_d$, so that the curve C2 is continuous
			       with C0.
  This is seen in the cooling diagram, presented in Fig \ref{cool},
where the number density of hydrogen in the halo, $n ({\rm cm}^{-3})=
\cM_{10} / (130 x_v^3)$, is plotted against the virial temperature,
   $T_v$ and shows the curves given by C0, the lower curve, C1, the
 vertical line, and C2 which is the upper curve in the region to the
    left of C1. The cosmological parameters were set to $h_{70}=1,
\Omega=1$, and $f_b=0.1$ and the halo parameters were chosen to be
 $\lambda_v=0.075, f_\ast=0.2$, and $\nu_{200}=4$. The shaded region
indicates the halos that have collapsed but star formation has induced
 gas loss and they ultimately resulted in gas poor systems. The halos
 above this region retain the gas that could form the massive central
			     black holes.

\begin{figure}
\resizebox{\hsize}{!}{\includegraphics{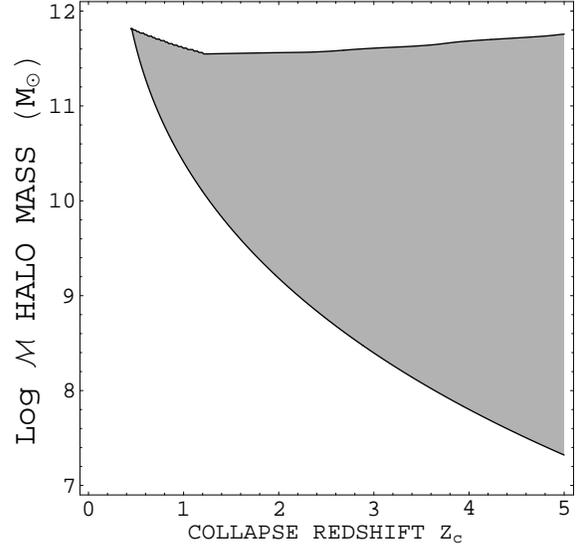}}
\caption{The halos in the shaded region in the $\cM_{10}-z_c$ space 
can trap the gas; the parameters chosen are $f_\ast=0.3, \nu_{200}=4$,
		 and $\lambda_v=0.07$.} \label{m-z}
\end{figure}

 The corresponding range in mass for a given redshift is shown in Fig
\ref{m-z}, for the choices of the parameters, $\nu_{200}=4$, the
  supernova efficiency, and $f_\ast=0.2$, the fraction of baryons in
stars.  Clearly for large redshift, the mass range increases and this
 is due to the deeper potential well which can retain the gas better.
 There is a crucial question of whether all the gas is retained or if
some fraction is lost in a wind. A supernova efficiency in the halo of
\begin{equation}
\nu_{200} \, {4 f_\ast \over 1- f_\ast} \gsim 60 \sqrt{\cM_{10}/x_v^3}= 4.3 
			    (1+z_c)^{3/2},
\end{equation}
from eqn(\ref{c2}), would be required to drive winds that would result
  in mass loss and this is considerably more than the corresponding
	     estimates in our galaxy, for $z_c \gsim 2$.

 If the stars form after the disk forms, the cooling rate in the disk
			  would be given by
\begin{equation}
\dot{E}_d= 3 \times 10^{44}~\Lambda_{23} \left ({f_b \over 0.1} \right 
  )^2\cM_{10}^2 ~x_d^{-3} (1-f_\ast)^2 {r_d \over H}{\rm ergs~
			       s}^{-1}
\end{equation}
     and the cooling time would reduce by a factor of a thousand
  corresponding to enhanced density.  Although only the cosmological
abundance of $[Fe/H]=-4$ was used, the supernovae explosions will also
enhance line cooling from metals injected into the medium; this would
	       reduce the cooling time estimated here.

\section{Magnetization of the disk}
\label{bdisk}
 In the large number of the halos which trap the gas as given by the
conditions C0--C2, a gaseous disk forms with a radial extent given in
\S \ref{collapse}.  Further, it was seen in \S \ref{shell} that the
supernovae shells fill the core volume at the time of star formation,
  and the small scale magnetic fields are dragged with the gas as it
settles into a the disk.  Now, we consider the question of whether the
   field strength is large enough to provide a significant viscous
 stress.  We take the gas to be initially dominated by the gravity of
 the dark halo and assume the following logarithmic potential (Binney
\& Tremaine 1987, eqn (2-54)) that obeys the flat rotation curve
\begin{equation}
\Phi(r,z)= v_c^2~ \ln(r_0^2+r^2+(z/q_\phi)^2) 
\end{equation}
  where $r_0$ is the radius of a compact region, $q_\phi \leq 1$ and
$v_c$ is defined in eqn (\ref{esc}).  If the vertical equilibrium was
 a result of balance between vertical gradient in the total pressure,
				 $P=
\zeta c_s^2 
\rho$ (which represents a sum of magnetic and gas pressures and the density 
	 scale height is $H$), and $-\del_z \Phi$, we obtain
\begin{equation}
  {H \over r} \simeq \sqrt{\zeta} \left ( { c_s \over v_c} \right )
			      q_\phi^2.
\label{vertical}
\end{equation}
Assuming a dominant isothermal halo, we obtain the half-flaring angle
			     of the disk,
\begin{equation}
		{H \over r_d} \simeq \theta~x_v^{1/2}
\cM_{10}^{-1/2},
\label{h} 
\end{equation}
 where $\theta =(1/8) ~\sqrt{\zeta} ~ (q_\phi^2 /0.5)$, and the speed
      of sound was taken to be $c_s= \sqrt{ {\gamma k_B T \over
\mu 
  m_p}} =1.6 \times 10^6$ cm s$^{-1}$ for a temperature of $10^4 K$,
	     where the cooling curve drops significantly.

   The supernovae inject the medium with magnetic flux and next we
     estimate the typical strength of the small scale field.  The
  calculation in \S \ref{shell} shows that the filling factor of the
  shells in the total gas volume is nearly unity for a wide range in
density and therefore the supernovae hot gas fills the medium and
  hence quasi-uniformly magnetizes it. The volume of the flared disk
			   works out to be
\begin{eqnarray}
  V_d &=& {4 \pi \over 3} ~r_d^2~ H \nonumber \\&=& 1.67 \times 10^8~{\rm pc}^3(8 \,
\theta)\left ({\lambda_v \over 0.075} \right )^3 \left({4.6 \over
		  1+z_c } \right )^3 \cM_{10}^{1/2}
\end{eqnarray}
  The Crab Nebula of size 0.8 pc (or volume 2 pc$^3$) has $3 \times
  10^{-4}$G fields (see \S \ref{disc} for a discussion of supernova
field strengths). By freezing the flux in a Crab volume of 2 pc$^3$ to
  a volume occupied by one shell, the magnetic field strength in the
disk for $N =10^6 \cM_{10} \nu_{200} (f_\ast/0.2) (f_b/0.1)$ remnants
				  is
\begin{eqnarray}
  B_{sn} &=& 3 \times 10^{-4} ~G ~\left ({V_d/N\over 2 ~{\rm pc}^3}
\right)^{-2/3} \nonumber \\
     &=& 1.7 \times 10^{-5} G ~\left({1+z_c \over 4.6} \right )^2
      ~\cM_{10}^{1/3}\nonumber \\&&\times \nu_{200}^{2/3}~\left ( {f_\ast \over 0.2}
\right)^{2/3}~ \left ( {f_b \over 0.1} \right)^{2/3}~ \left ({0.075
\over \lambda_v} \right)^2.
\label{bsn}
\end{eqnarray}
  By choosing a typical set of values ($\lambda_v=0.05, \nu_{200}=4,
			     f_\ast=0.5,
\theta=1/8$) the field strength turns out to be as high as $B_{sn}=170 \mu G$.
 
The mean number density of hydrogen atoms in the flared gaseous disk,
		      for $\mu=0.6$, is given by
\begin{eqnarray}
  n_d &=& 7.5 ~(1-f_\ast) ~M_9~ \left ({r_d \over H} \right)~ x_d^{-3}\nonumber \\
 & =&150 ~{\rm cm}^{-3}~ {1-f_\ast \over 0.8} ~\left({1+z_c \over 4.6}
 	    \right )^{7/2} ~\cM_{10}^{2/3}~{f_b \over 0.1}
\label{nd}  
\end{eqnarray}
     where (\ref{h}) and (\ref{gas}) were used and the values of
$\lambda_v=0.075$ and $\theta =1/8$ were taken. Outflows from O and B
  stars could also magnetize the gas, and by using flux freezing we
estimate the field strength due to winds to be (Bisnovatyi-Kogan et al
	     1973; Ruzmaikin, Shukurov \& Sokoloff 1988)
\begin{equation}
  B_{w}=(\rho/\rho_w)^{2/3} B_s = 5.2 \mu G ~\left ({n_d \over 150}
\right)^{2/3}.
\label{wind}
\end{equation}
 where the estimates of the density at the base of the wind, $\rho_w
\simeq 2
\times 
 10^{-13}$ g cm$^{-3}$ and the field at the surface of the star, $B_s
\simeq 4 G$ are used. The field expelled by massive stars are unlikely
   to pervade the volume, and the major contribution would be from
supernovae. Although the number of massive stars are of the same order
   as the number of supernovae using the Salpeter IMF (estimated by
		 integrating the mass range above $10
\Msun$; 
			 see the footnote in
\S 
\ref{halo}), the smaller fluxes of the wind ($\sim B_s R_\ast^2$ where $R_\ast$ 
 is the radius of the star) render the effective field strength to be
weak. {\em The key point here, is that for typical values of the halo
parameters, $B_{sn}$ is about $10^{-4}-10^{-5}$ G which is a factor of
  10--100 higher than the value in our galaxy ($V_d=6 \times 10^{11}
			    {\rm pc}^3, N
\sim 3 \times 10^8, B_{sn} = 3 \mu G$) due to a smaller  value of  $V_d/N$ of 
		       the supernovae shells}.

\section{Accretion to a compact region}
\label{viscosity}
  Here we consider different viscosity prescriptions, namely direct
 magnetic stress, the phenomenological $\alpha$ viscosity (Shakura \&
   Sunyaev 1973) derived from magnetic fields, and angular momentum
transfer mediated by self-gravity induced instabilities, and calculate
   the corresponding accretion timescales. The time dependent disk
 accretion is described by the conservation of mass, radial momentum
(where all other forces except gravity are neglected, and $v_r/v_\phi
\ll 1)$ and angular momentum
\begin{eqnarray}
   r\partial_t \Sigma&=&- \partial_r(r \Sigma~ v_r) \label{mass}\\
\omega^2 &=& (1/r) \partial_r \Phi \\
\Sigma v_r \partial_r(r^2 \omega)&=& 1/r ~\partial_r (r^2 \Pi_{r\phi}), 
\label{disk}
\end{eqnarray}
      where $\Sigma(r,t)$ is the surface density, $\Phi$ is the
gravitational potential and $\Pi_{r\phi}$ is the vertically integrated
  stress. We take $r_0$ as the inner radius of the disk flow and the
		outer edge of the compact region while
\begin{equation}
\omega(r) = {\omega_0 \over \sqrt{1+(r/r_0)^2}}.
\end{equation}
      By integrating eqn (\ref{disk}), we obtain for steady flow
\begin{equation}
\dot{M} \omega \left (1-{r_0^2 \omega_0 \over 2 r^2 \omega} \right )= 2 \pi 
\Pi_{r\phi}.
\label{mdot}
\end{equation}
 The viability of the various alternatives for the viscous stress can
be assessed by estimating the accretion rates. In the initial phase of
 accretion the potential is dominated by the halo as calculated in \S
\ref{bgd}, while in the later phase, self-gravity dominates and the
	    resulting flow is calculated in \S \ref{self}.
 
\subsection{Magnetic Stress}
\label{magnetic}
We consider now the form of the stress tensor that is entirely due to
   magnetic fields.  We assume that the magnetic field injected is
  largely in the form of small scale loops that are frozen into the
   plasma. Further, we expect that the processes of compression and
 		advection preserves the form given by
\begin{equation}
		      B= B_s (\rho/\rho_s)^{2/3}
\label{fluxf}
\end{equation}
       that is likely to be valid for small scale fields in the
      non-dissipative limit.  The Lorentz force on the plasma is
\begin{equation}
 ({\bf J} \times {\bf B})_\phi ={1\over 4 \pi} ~ \left [{1 \over r}\,
				 B_r
\,\partial_r (r B_\phi) +  B_z \,\partial_z B_\phi \right ],
\end{equation}
where the first term on the right hand side is the local shear stress
 and the second term is negligible in the initial phase (the vertical
	     average of small scale fields, $B_z \langle
\partial_z B_\phi \rangle 
\approx 0$), but it could be important during a later phase of accretion 
where magnetic braking can operate via built up, large scale $B_z$, by
    angular momentum transfer to the parts external to the compact
  region. Here, we take only the local shear stress $\approx B^2 H/
  (2\pi)$ and this was assumed to be initially at sub-equipartition
levels as a thermal pressure ($\zeta=1$) was used for calculating the
 half-thickness of the initial disk in eqn (\ref{h}). One can verify
 this through a consistency check by using the definition, $\zeta= 1+
B_{sn}^2/(8 \pi \rho c_s^2)$, which determines the half-thickness, $H
\propto
\sqrt{\zeta}$ from the condition of vertical equilibrium, (\ref{h}), $B_{sn} 
\propto \zeta^{-2/3}$, from eqn (\ref{bsn}), and $\rho \propto \zeta^{-1/2}$ 
		 from eqn (\ref{nd}). It follows that
\begin{equation}
0.		    03 \, \zeta^{-1/6} + 1= \zeta
\end{equation}
 and for a reasonable choice of parameters ($\nu_{200}=1, f_\ast=0.2,
			       z_c=3.6,
\lambda_v=0.075$) this results in $\zeta =1.03$. Low values of 
 $1/\beta=B_{sn}^2/(8 \pi \rho c_s^2) \sim 0.03$, are typical for the
	      range of interest in the parameter space.

Next, we estimate the accretion time scale and hence the viability of
  magnetic accretion in the steady limit.  For typical values of the
	     parameters assumed here ($\lambda_v = 0.075,
\theta=1/8$) 
		  the accretion rate turns out to be
\begin{eqnarray}
\dot{M}&=& \left |{2 \pi \Pi_{r\phi}^m \over \omega} \right |\sim{B^2 H \over 
\omega}= 
\left ({ H \over 
r_d} \right) {r_d^2 B^2 \over v_c}\nonumber \\ &=& 0.2 ~B_{-5}^2 ~\left( {4.6 \over
		1+z_c} \right)^3~\Msun~{\rm yr}^{-1}.
\end{eqnarray}
using eqns (\ref{h}, \ref{esc}, \ref{mdot}). Taking the initial field
  to be from supernovae expulsions ($B=B_{sn}$) from eqn(\ref{bsn}),
	   this implies a time scale of magnetic accretion,
\begin{eqnarray}
  t_m&=& 10^8 \, \Msun \,{\bhs \over \dot{M}} \nonumber \\
&\sim& 2 \times 10^8 ~{\rm
			      yrs}~ {4.6
\over 
  1+z_c}~\cM_{10}^{-2/3}\nonumber \\&&~\times\nu_{200}^{-4/3}~\left ({0.2 \over f_\ast}
\right )^{4/3}\,
\bhs\,\left ({0.1 \over f_b} \right )^{4/3}.
\label{tm}
\end{eqnarray}
Having obtained this fast timescale to accrete $10^8 \Msun$ of gas, we
   proceed to calculate the detailed form of $\Pi^m_{r \phi}$.  The
 thickness of the disk is given by eqn (\ref{h}), the balance of the
   total pressure gradient and vertical gradient of the background
       potential.  Combining eqns (\ref{h}), (\ref{fluxf}), and
		    $\rho=\Sigma/(2 H)$, we obtain
\begin{equation}
\Pi^{m}_{r\phi}= -(B_s^2/4 \pi) \rho_s^{-4/3} \Sigma^{4/3}  r^{-1/3} 
\theta_0^{-1/3}
\end{equation}
where $\theta_0= 2 \theta (\cM_{10} /x_v)^{1/2}$ is the flaring angle
of the full thickness of the disk. Initially, the magnetic pressure is
lower than the thermal pressure but as the matter sinks into a compact
  region the magnetic pressure is expected to dominate the vertical
			  pressure gradient.

\subsection{ $\alpha$ viscosity}
\label{alpha}
 If a small scale dynamo operates quickly (Kasantsev 1967, Kulsrud \&
Anderson 1992) then it will saturate near equipartition values (as is
   well known from simulations - Hawley, Gammie \& Balbus 1995 and
   references therein) and an appropriate form of the stress can be
described in terms of a prescription of the form $\Pi^{ss}_{r\phi}= -2
 H \alpha_{ss} P$, where $P= \zeta \rho c_s^2$ is the total pressure,
   which is proportional to the gas pressure at equipartition. The
  accretion time scale is expected to be similar to the one obtained
 earlier.  The dependence on the halo parameters can be expressed in
		       the isothermal limit as
\begin{eqnarray}
t_\alpha &\sim& 2 \times 10^8 ~{\rm yrs}~ {0.1 \over \alpha_{ss}}~\left
				({4.6
\over 
1+z_c} \right)^{1/2}~{0.1 \over f_b}\nonumber \\
&&\times~\cM_{10}^{-2/3}~\left ({0.8 \over
			      1-f_\ast}
\right ) \, \bhs.
\label{talpha}
\end{eqnarray}
      Note that, although we use a direct magnetic stress in our
   calculations, we record for comparison, the detailed form of the
	     $\alpha$ prescription in Appendix \ref{clg}.

\subsection{Gravitational instabilities}
\label{grav}
   Cold, thin rotating discs are known to be unstable and the basic
  stability criteria was provided by Toomre (1964). For a uniformly
		 rotating isothermal disk (Goldreich
\& Lynden-Bell 1965) the criteria for local stability is given by
\begin{equation}
	     Q_T ={c_s \omega \over G \Sigma} \geq 1.06.
\end{equation} 
			     We find that
\begin{equation}
Q_T= {\pi v_c^2 H \over G M_d} = {\pi \over 8}~ \cM_{10}^{-1/3}~\left
				({4.6
\over 
      1+z_c} \right)^{1/2}~{0.8 \over 1-f_\ast}~{0.1 \over f_b},
\label{toomre}
\end{equation}
where the values ($\theta=1/8, \lambda_v=0.075$) were assumed. So the
disk is unstable to gravitational instabilities and it is possible for
   angular momentum transport to occur through this process. Lin \&
     Pringle (1987) estimate an effective kinematic viscosity for
   gravitational instability from $\nu_e=\omega \ell^2$, where the
critical shearing length, $\ell=G \Sigma \omega^{-2}$ was taken to be
  the maximum possible size for the instability. Here we make a more
    conservative estimate by introducing the parameter, $Q_T^{2}<
\alpha_g<1$, into the stress given by
\begin{equation}
\Pi^g_{r\phi}=\alpha_g ~\nu_e ~\Sigma \,r \partial_r \omega =-\alpha_g~G^2\, 
\Sigma^3\, \omega^{-2} ~~~{\rm if} ~~Q_T <1.
\label{pig}
\end{equation}

	     The corresponding timescale of accretion is
\begin{eqnarray}
      t_g &\sim& 10^8 \, \Msun \, \left | {\omega \,\bhs \over 2\pi
\,\Pi_{r\phi}^g}\right 
| \\&=& 10^8 {\rm yrs}\cM_{10}^{1/2}\left ({4.6 \over 1+z_c}
\right)^3\left ({0.8 \over 1-f_\ast}\right)^3
\left({0.1 \over f_b}\right)^3 \, {0.1 \over \alpha_g} \,\bhs. \nonumber
\label{tg}
\end{eqnarray}
\section{Self-similar evolution of the disk in a background potential}
\label{bgd}
Having demonstrated that the time scales of accretion are quite fast,
we now proceed to calculate a detailed model of self-similar evolution
 of a disk from the diffusion equation obtained from the conservation
		     laws (\ref{mass}-\ref{disk})
\begin{equation}
\del_t \Sigma = {1 \over r} \partial_r \left ({1 \over \del_r(r^2 \omega)} 
\del_r[r^2 
\Pi_{r 
\phi}] \right),
\label{diff}
\end{equation}
 where the viscous stress can be parameterized as $\Pi_{r \phi}=-K_2
\Sigma^b r^c$ and in addition the rotation law is assumed to be of the
form $\omega=K_1 r^a$. This very useful formulation of a self-similar
 form is due to Pringle (1981) but only particular analytic solutions
to the diffusion equation has been reported for the specific cases of
($a=-3/2, b=c=3$; Lin \& Pringle 1987) in the context of accretion of
a protostellar disk onto a point mass via gravitational instabilities
and ($a=-3/2, b=5/3, c=-1/2$; Cannizzo, Lee \& Goodman 1990 (CLG), see
  Appendix \ref{clg} of this paper) in the context of disk accretion of a tidally
   disrupted star onto a massive black hole. Note that in CLG, the
scaling law for the viscous stress by the closure of the conditions of
local dissipation in an alpha disk in a Kepler potential and vertical
equilibrium. Here, an analytic solution to the general problem of the
	 type ($ \Pi_{r\phi} \propto \Sigma^b \, r^c, \omega
\propto 
  r^a$) is presented so that possible viscosity mechanisms discussed
 earlier and expressible in this way, can be explored within the same
 formulation. In the magnetic case given below, the viscosity scaling
  is due a magnetic stress, the flux-freezing condition and vertical
    equilibrium in a cold disk in the background halo potential. A
   solution for an alpha disk with local dissipation with a general
 rotation law is provided in Appendix \ref{clg}. The general solution
   presented below has a larger utility in contexts other than one
			   considered here.

\begin{figure}
\resizebox{\hsize}{!}{\includegraphics{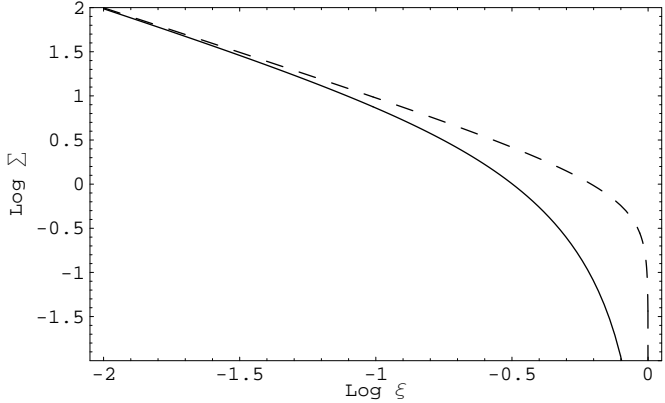}}
\caption{The structure $g(\xi)$ of the 
	 self-similar disk, $\Sigma=\Sigma_0 \,\left({t \over
 t_0}\right)^{\beta} \,g(\xi)$ where $\xi=r/r_f, r_f= r_s \,\left({t
\over t_0}\right)^{\alpha}$. The magnetic solution,
   $g_m(\xi)=\xi^{-1/2} \left [1-\xi^{3/2}\right]^3$ is shown by a
solidline, and the solution of the disk with gravitational viscosity,
			  $g_g(\xi)=\xi^{-1}
\left [1-\xi\right]^{1/2}$, is shown by a dashed line.}
\label{gxi}
\end{figure}

 If $b=1$, the equation is linear and the general solution is easily
found. Proceeding generally, under the assumptions of self-similarity
    for ($b \neq 1$), one may write the the surface density in the
			    following form
\begin{eqnarray}
\Sigma =\Sigma_0 (t/t_0)^\beta g(\xi) && r_f= r_s (t/t_0)^\alpha,
\label{dense}
\end{eqnarray}
where $\xi \equiv r/r_f$, and $r_s$ is the associated radius scale. We
			  set the constants
\begin{equation}
 t_0= {K_1 \over K_2} r_s^{a+2-c} \Sigma_0, ~M_d=2 \pi l \Sigma_0
	     r_s^2, ~l=\int_0^1 \xi^2 g(\xi) \diff \xi
\end{equation} 
   where $M_d$ is the initial disk mass. Here we seek a particular
  solution when there is no external torque, which implies the total
 angular momentum, $J$, of the disk is a constant. Using the scaling
      relations above that are implicit in eqn (\ref{diff}) and
\begin{equation}
J= 2 \pi K_1 r_s^{4+a} \Sigma_0 j, ~~~~~j= \int_0^1 \xi^3 g(\xi) \diff
\xi
\end{equation}
    it follows that $\alpha=(4 b +ab-2-c)^{-1}$ and $\beta= -(a+4)
\alpha$. At this point we note that the disk edge travels outward if
$2+c < b (4+a)$.  Substituting into the form for the surface density,
as given in (\ref{dense}), and simplifying (\ref{diff}), we obtain the
	      following ordinary differential equation,
\begin{equation}
-  \alpha \xi g' + \beta g= -{1 \over (a+2) \xi} \diff_\xi \left (
 			     \xi^{-(a+1)}
\diff_\xi 
		      (\xi^{2+c} g^b) \right ).
\end{equation}
  After some algebraic transformations, one can integrate it once to
				obtain
\begin{equation}
\alpha (a+2) g \xi^{a+4} +(a-c) \xi^{2+c} g^b = b \xi^{3+c} g^{b-1} g' +c_1.
\end{equation}
Now, we apply the boundary condition that the density vanishes at the
			   disk edge, ie.,
\begin{eqnarray}
\Sigma(r=r_f)=0; ~~g(1)=0.
\end{eqnarray}
  Moreover, if $b >1$, which is the case for the examples considered
 here, then $c_1=0$.  By rearranging terms and integrating, we obtain
			the following solution
\begin{equation}
  g(\xi)=\xi^{(a-c)/b} \left (1-\xi^{2+{(a-c) \over b}} \right)^{1 /
			       (b-1)}.
\label{solution}
\end{equation}
       The time constant can be evaluated with $r_s=r_d$ to be
\begin{equation}
 t_0 = {\omega \over \left |\Pi_{r\phi}(\Sigma_0, r_d) \right|} {M_d
\over 2 \pi l}=t_a l^{b-1}~{M_{d8} \over \bhs},
\end{equation}
       where $t_a$ is the accretion time scale calculated in \S
\ref{magnetic} and \S
\ref{grav} in the steady case using $\Pi_{r\phi}(M_d/(2 \pi r_d^2), r_d)$.
       The rate at which mass sinks into the center is given by
\begin{equation}
   M_c(t)= M_d \left [1- \left ({t \over t_0}\right)^{-(a+2)\alpha}
\right]
\label{sink}
\end{equation}
  The accretion time scale which is the time that transpires when a
 fraction $\epsilon=\bhs/M_{d8}$ of the disk mass falls in is given by
\begin{eqnarray}
\tau_a&=& t_0~\left ((1-\epsilon)^{-1 \over (a+2) \alpha} -1\right) \nonumber 
\\&=& t_a  ~l^{b-1}~\left ((1-\epsilon)^{-1 
\over (a+2) \alpha} -1\right)~\epsilon^{-1}
\end{eqnarray}
 where $t_a$ represents the timescales $t_m, t_g$ or $t_\alpha$, and
indicates that the estimates derived earlier are modified by geometric
 factors.  Now we consider the particular case of magnetic accretion
			($a=-1, b=4/3, c=-1/3,\epsilon=1/8$,
 see \S \ref{magnetic} where the value of $\epsilon$ is chosen for a
 typical case where the disk mass, $M_{d8}=8$, and $\bhs=1$) which has
			     the solution
\begin{eqnarray}
\alpha=3/7, &&~\beta=-9/7, ~\tau_m= 1.01 \, t_m, \nonumber \\
 &&~ g_m(\xi)=\xi^{-1/2} \left [1-\xi^{3/2}\right]^3.
\label{magsoln}
\end{eqnarray}
 Similarly, the accretion due to gravitational instabilities ($a=-1,b=3, c=2,
\epsilon=1/8$, $M_{d8}=8$, and $\bhs=1$, see \S 
\ref{grav}) has the 
			       solution
\begin{eqnarray}
\alpha=1/5,&&~\beta=-3/5, ~\tau_g= 0.54 \,t_g,\nonumber \\ 
	   && ~g_g(\xi)=\xi^{-1} \left [1-\xi \right]^{1/2}.
\label{gravsoln}
\end{eqnarray}
  The disk structure of these solutions are shown in Fig. \ref{gxi}.

 The self-similar solutions of this kind to (\ref{diff}) are known to
  develop at large times in numerical simulations with a variety of
	      initial conditions (Lin \& Pringle 1981).

\begin{figure}
\resizebox{\hsize}{!}{\includegraphics{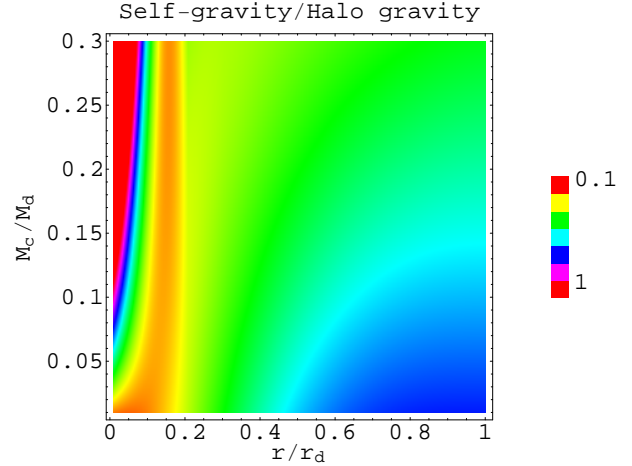}}
\caption{A density plot of the ratio of 
 self-gravity (disk and the central mass) to halo gravity defined as
   ${1 \over v_c^2} \left (r {\diff \Phi_d(r,t) \over \diff r} + {G
 M_c(t) \over r} \right)$, showing the evolution of a magnetized disk
	in the halo potential as the central mass increases to
 $\epsilon=M_c/M_d=0.3$.  The horizontal axis is the in units of the
   initial disk radius, $r_d$. The value of $\lambda_v (1-f_g)/(f_g
\sqrt{k_1})$ was chosen to be 3; the halo gravity (in units of
$v_c^2$) dominates at radii outside a central region of $y \sim 0.1$.}
\label{mbh}
\end{figure}

    Now we determine the regime in parameter space where the halo
   dominated flow can occur. This is given by the condition that at
       $t=t_0$ and $r=r_d$, the halo dominates self-gravity or
\begin{eqnarray}
   {v_c^2 \over r_d}&>& \left. {\diff \Phi_d \over \diff r} \right
|			       _{r=r_d}
\nonumber 
\\
   {M_h \over r_d r_v} &>& {M_d \over l r_d^2} ~ \sum_n \alpha_n^2
			    J_1(\alpha_n)
\,
\int_0^1 J_0(\xi \alpha_n) g(\xi) \xi \diff \xi, 
\end{eqnarray}
 where $v_c$ is the circular velocity due to the halo alone, the disk
 potential was expressed as a Bessel transform of the surface density
and $\alpha_n$ are the zeros of $J_0$. The collapse factor $r_v/r_d =
		    s \sqrt{k_1}/\lambda_v= (j/l)
\sqrt{k_1}/\lambda_v$ from \S \ref{collapse}, $M_h/M_d= (1-f_g)/f_g$ by 
			definition, leading to
\begin{equation}
   {\lambda_v (1-f_g) \over f_g \sqrt{k_1}} > {j \over l^2} \sum_n
\alpha_n^2 J_1(\alpha_n) \,
\int_0^1 J_0(\xi \alpha_n) g(\xi) \xi \diff \xi
\end{equation}
 For the solution (\ref{magsoln}), the magnetic case, the RHS of the
    above equation works out to be 0.44 and for the gravitational
  instability case, the solution (\ref{gravsoln}), the corresponding
	value is 0.46. Hence this condition can be written as
\begin{equation}
		 {\lambda_v (1-f_g) \over f_g} > 0.32
\label{gravity}
\end{equation}  
  where $k_1=0.5$ for a truncated isothermal sphere was used. For a
       reasonable range in the parameters ($\lambda_v=0.05-0.1,
 f_g=0.05-0.08$) the above condition holds good ($\lambda_v (1-f_g)/
f_g$ is in the range 1--4); as a result, the initial accretion flow is
 expected to be halo dominated. As the mass accretes into the center,
      the spin deviates from $\omega \propto 1/r$ and gradually
  increases. The self-similar solutions are valid only up to a point
   beyond which the self-gravity due the disk and the central mass
dominates the potential.  The time of transition to a self-gravitating
		flow can be estimated by seeking that
\begin{equation}
   r {\diff \Phi_d(r,t) \over \diff r} + {G M_c(t) \over r} > v_c^2
\end{equation} 
    Now expressing time in terms of the $\epsilon=M_c/M_d$ in eqn
	      (\ref{sink}), we obtain after some algebra
\begin{displaymath}
  (1-\epsilon)^3 {j y \over l^2}
\left [
\sum_n 
\alpha_n^2 J_1(\alpha_n (1-\epsilon)y)\, \int_0^1 J_0(\xi \alpha_n) g(\xi) \xi 
\diff \xi \right] \nonumber
\end{displaymath}
\begin{equation}
~~~~~+{\epsilon j \over l y (1-\epsilon)} > {\lambda_v (1-f_g) \over f_g \sqrt{k_1}},
\label{check}
\end{equation}
  where $y=r/r_d$. Clearly, this condition does not hold close to a
 compact central region (defined by $r<r_0$) where it is dominated by
      the central mass. The solutions considered here are a good
 approximation for the region beyond $y=0.1$ for $\epsilon < 0.3$ and
	provide a sufficiently accurate description (See Fig.
\ref{mbh}). 
   At large times, a Keplerian flow into the compact region can be
   assumed to occur and one can use a self-similar flow again with
$a=-3/2$ and the corresponding magnetic stress taking into account $v_c
\propto r^{a+1}$ in  eqn (\ref{h}) (in combination with (\ref{fluxf}), and 
  $\rho=\Sigma/(2 H)$) leads to $\Pi_{r\phi}^m \propto \Sigma^{4/3}
   r^{a/3}=\Sigma^{4/3} r^{-1/2}$. Similarly $\Pi_{r\phi}^g \propto
\Sigma^3 r^{-2a}=\Sigma^3 r^3$ from eqn (\ref{pig}). However, the
    estimate of the accretion timescale is not expected to be very
different from the one derived earlier. The key result is that about a
  fraction, 0.3, of the disk mass can be transported into a central
  region, which is a fraction, $y= 0.1$, of the initial disk radius
 within the time given by the solutions (\ref{magsoln}). For typical
  values, this implies that about $2 \times 10^8 \Msun$ of gas sinks
into a $r_0= 0.1 r_d \sim 100$ pc region in a halo (with $\cM_{10}=1,
\lambda_v=0.075$) in a timescale of $2 \times 10^8$ yrs.
\section{Disk evolution in a self-gravitating regime}
\label{self}
 We now examine the evolution of the disk in a gravity field that is
entirely due to itself. This flow can occur after sufficient accretion
 of mass into a compact region of radius $r_0$; or, alternatively at
 the time of the formation of the disk if the eqn (\ref{gravity}) is
not satisfied and $\lambda_v (1-f_g)/ f_g \lsim 0.32$. The problem of
  self-gravitating accretion flow is complicated by the coupling of
Poisson's equation to the momentum and continuity equations. Clearly,
its evolution has to be treated differently from the preceding case of
 a prescribed background potential. A detailed model is deferred to a
   work in preparation (Mangalam 2001); below, we consider a useful
simplified version of the problem by assuming a particular form of the
			density distribution (see Field 1994).

   We assume a Mestel (1963) disk where the self-consistent density
distribution with potential is entirely due to self-gravity, is of the
				 form
\begin{equation}
\Sigma(r,t)= {v_\phi^2(t) \over 2\pi G r},
\end{equation}
       where the time dependence appears only in the rotational
  velocity. This is a an interesting disk which is self-consistent,
	taking into account the most relevant physics. Taking
\begin{equation}
    v_\phi= v_0 \,\chi(t) ~~~{\rm and}~~~ r=r_0 \, \chi_1(t) \, x
\end{equation}
 where $v_0^2= G M_c/r_0$, and $M_c$ is the mass out to $r_0$. We see
that by assuming a self-similar evolution of the disk, the mass out to
  a given $x$ should be independent of $t$ and hence it follows that
\begin{equation}
\chi_1 = \chi^{-2} ~~~ {\rm and} ~~ \Sigma= \Sigma_c \,{\chi^4 \over x},
\end{equation}
 where $\Sigma_c= M_c/ (2\pi r_0^2)$.  From the continuity equation,
\begin{equation}
	       r\del_t \Sigma + \del_r(r \Sigma v_r)=0
\end{equation} 
			       we find
\begin{equation}
		 v_r= -4 \chi^{-3} \dot{\chi} r_0 x.
\end{equation}
  Substituting this and the self-similar forms given above into the
		      angular momentum equation,
\begin{eqnarray}
\Sigma \del_t v_\phi +\Sigma (v_r/r) \del_r(r v_\phi)&=& (1/r^2)\del_r(r^2 
\Pi_{r\phi}) \nonumber \\
-3 v_0 \Sigma_c \chi^2 \dot{\chi} &=& {1 \over x}{\diff \over \diff x} \left (x^2
\Pi_{r\phi}\right),
\end{eqnarray}
			      we obtain
\begin{equation}
\Pi_{r\phi}=-{3 \over 2} v_0 \Sigma_c r_0 \chi^2 \dot{\chi},
\label{chi}
\end{equation}
   which is {\em independent} of $x$. So far no specific viscosity
   mechanism has been invoked-- the form of $\Pi_{r\phi}$ above is
   necessitated by the prescription of a Mestel disk. If a magnetic
  stress is assumed and $\Pi_{r\phi}= -\kappa B^2 H/(2 \pi)$, where
 $\kappa$ is a factor of order unity and $B$ is given by the vertical
balance of magnetic pressure and gravity, $B=2 \pi \sqrt{G} \Sigma= 2
\pi 
\sqrt{G} \Sigma_c \chi^4/x$. It follows that the half thickness, $H \propto x^2$, and  can be expressed as 
$H= H_0 x^2 \chi^p$. Furthermore, from the flux-freezing condition, $B
\propto
\left 
[r^2 H \right]_{x=1}^{2/3}$, it is seen that $p=-2$. Equating the form
of $\Pi_{r\phi}$ from eqn(\ref{chi}) to a magnetic stress, $\kappa B^2
				H/ (2
\pi)$, 
       and writing (after taking $\theta=1/8, \lambda_v=0.075$)
\begin{eqnarray}
\tau&=& {1 \over 2} {v_0 r_0^3 \over G M_c H} ={r_0 \over v_0}~ {1 \over 2 \kappa} 
  ~ {r_0 \over H_0} \nonumber \\
&\sim& 10^6 ~\left ({r_0/r_d \over 0.1} \right) ~\cM_{10}^{1/3} ~ {4.8
\over  1+z_c} ~{\rm yrs},
\end{eqnarray}
			     we see that
\begin{equation}
		 3 \tau ~\chi^2~ \dot{\chi}= \chi^6,
\end{equation}
		     which leads to the solution
\begin{equation}
\chi(t)=  \left ( {\tau \over \tau -t} \right)^{1/3},
\end{equation}
  where $\chi(0)=1$ was taken as the initial condition. So the disk
    spins up rapidly and shrinks to a smaller radius. Clearly, the
solution is no longer valid when it is relativistic. The self-similar
 collapse of the compact region is only a sketch but nevertheless the
collapse timescale, $\tau$, suggests that formation of a black hole is
		  extremely rapid ($\sim 10^6$ yrs).

\section{Summary of the results and discussion}
\label{disc}
		 We summarize our results as follows:
\begin{enumerate}
\item 
We considered star formation and supernovae feedback on the remaining
 gas in the halo based on the framework of DS.  We include a new and
  necessary condition C2, namely that escape time for the hot gas be
shorter than the cooling time and find that the condition for gas loss
   can be expressed by eqns (\ref{c0}, \ref{c1}, \ref{c2}) which is
   depicted in Fig \ref{cool}. For a typical choice of $f_\ast$ and
$\nu$, Fig \ref{m-z} shows the allowed range for Halo mass in terms of
    redshift. There is a sharp decline in the allowed range beyond
   collapse redshifts of $z_c=2$, as the potential wells formed at
	  earlier epochs are deeper and trap the gas better.

\item
 It was seen in \S \ref{shell} that the hot gas from supernovae has a
  filling factor of nearly unity and hence the small-scale magnetic
  fields occupy the disk.  It was shown in \S \ref{bdisk} the field
  strength is significant ($10-100\mu$G), based on typical values of
      supernova efficiency, $\nu$, and star fraction, $f_\ast$.

\item
 In \S \ref{viscosity}, we examined magnetic stress and viscosity due
    to self-gravity induced instabilities that would operate in a
 collapsed disk. The accretion timescales were estimated for a direct
    magnetic stress, eqn(\ref{tm}), an $\alpha$ prescription, eqn
  (\ref{talpha}), and self-gravity, eqn(\ref{tg}), in terms of halo
	parameters $(\cM, z_c)$ and star formation parameters
 $(f_\ast,\nu)$. The timescales are all short compared to the cosmic
				time.

\item
  A general solution for a self-similar and time dependent accretion
 flow for a viscous stress of the form $\Pi_{r\phi} \propto \Sigma^b
r^c$ in a prescribed potential, $\omega \propto r^a$, was obtained in
\S \ref{bgd}. This was applied to the specific cases of magnetic
  accretion, eqn(\ref{magsoln}) and and gravitational instabilities,
 eqn(\ref{gravsoln}). The structure of the resulting disk is shown in
				 Fig.
\ref{gxi} and the timescales are within geometric factors of their estimates.  
  The disk eventually becomes fully Keplerian. The condition of halo
dominated flow is given by eqn(\ref{gravity}) and the transition to a
 self-gravitating flow is given by eqn(\ref{check}). This solution is
 valid for dominant halos in the initial stages for the outer parts,
  $y>0.1$, oustide a Keplerian compact region for up to time when a
  fraction of the disk mass, $\epsilon \sim 0.3$, falls in (see Fig
\ref{mbh}).

\item
A self-gravitating Mestel disk that is spinning up as it is collapsing
  self-similarly, is described in \S \ref{self}. We apply this to a
  compact region where the pressure is assumed to be due to magnetic
 fields. The time scale of collapse turns out to be the rotation time
	  of the outer radius, which is a few million years.
\end{enumerate}
	     We now discuss some of the issues involved:

\noindent{\em Supernovae feedback}

 In a pioneering work, DS, showed that with a model of protogalactic
  gas in a halo reproduces the observed relations very well with an
 initial CDM model.  Further, the condition for gas loss was given by
 the condition that the energy input from supernovae is more than the
 binding energy, C1 and the cooling condition for gas contraction and
   star formation. In this work, we specify an additional necessary
   condition, C2, that the escape time be shorter than the cooling
time. As indicated in Fig. \ref{cool}, this restricts the gas loss to
 the halos in the shaded region. The exact shape and location of the
region is subject to the choice of parameters ($\nu, f_\ast$), but the
key point made here is that the entire region to the left of C1 cannot
  be considered as a gas loss zone, even if heating by background UV
photons were included. It is worth investigating this question in more
 detail by simulations that include the hydrodynamics and cooling of
 the hot gas. Based on this study, one can conclude that most of the
    dwarf galaxies involved efficient conversion of stars or high
 efficiency of supernovae, since that would place curve C2, higher on
 the cooling diagram accounting for the prediction of the location of
the dwarf galaxies made by DS. The constraints imposed here, however,
account for the presence of black hole systems on the left of C1 (and
  above C2).  It also explains a sharp decline in quasars at epochs
			  later than $z=2$.
  
\noindent{\em Magnetic field strength in supernovae shells}

 In estimating the field strength $B_{sn}$ in \S \ref{bdisk}, all the
 shells were assumed to have $10^{-4}$ G when they are young (size of
0.8 pc) and flux freezing was used to estimate the values at a given
  size. This field has been measured in the Crab nebula and in other
   young supernova remnants like Tycho, Kepler and Cas A, the field
strength in the shell is inferred to be $10^{-3}-10^{-4}$ G (Strom \&
Duin 1973, Henbest 1980, Anderson et al. 1991).  Simulations by Jun \&
 Norman (1996), show local turbulent amplification by Rayleigh-Taylor
   and Helmholtz instabilities but claim overall sub-equipartition
strengths. An analysis of Cas A X-ray and radio surface brightness at
 high resolution, show that there is a strong correlation, implying a
 possible equipartition between the field ($\sim$ mG) and the hot gas
			 (Keohane, Gotthelf,
\& 
  Petre 1998). The field amplification can be from other sources as
well. For example, the magnetic field in the Crab has been wound up in
the body of the nebula by the pulsar and the magnetic energy is a few
  a percent of the spin energy (Rees \& Gunn 1974). High resolution
  simulations of a supernova including rotation and expanding in an
unmagnetized medium, which to our knowledge are unavailable, are best
suited to the answer the question of the initial field strength in the
disk. The small scale fields that are injected in the medium could be
amplified to saturation levels by a dynamo operating in the disk aided
      by differential rotation and turbulent motion. Since this
  amplification was not taken into account (no assumption other than
 flux-freezing was made for the field evolution), the field strengths
derived from supernovae which is in the range 10-100 $\mu G$, could be
			  an underestimate.

\noindent{\em Large scale fields}

Strong fields of a few $\mu G$ are estimated from Faraday rotation of
 background QSOs by damped Ly$\alpha$ absorptions systems, which are
 thought to be proto-galactic disks, as early as $z \sim 2$, imposing
  constraints on the kinematic dynamo (Kronberg, Perry, \& Zukowski
1992).  A primordial origin for the large scale field in galaxies has
 been speculated upon (Rees 1994, Ratra 1992 and references therein),
 but the field generated by these mechanisms is much smaller than the
 magnitude required for a primeveal hypothesis.  Large-scale magnetic
fields in the galaxy are thought to be generated by a turbulent dynamo
 from a weak seed field ($\sim 10^{-19} G$). But, it has been argued
that magnetic energy at small (eddy) scales builds up much faster (at
eddy turnover rates) than the mean field, so that the kinematic dynamo
  shuts off before the large scale fields amplify to observed levels
  (Kulsrud \& Anderson 1992, Kasantsev 1967) if the initial seed is
weaker than $10^{-9}$G.  Subramanian (1998) argues, however, that the
 fields are intermittent and do not fill the volume. The small scale
 flux ropes are at saturated values but the mean energy density is at
  sub-equipartition levels allowing the large scale field to grow to
  equipartition strength, aided by ambipolar drift. This question is
  still unresolved, but if the supernovae field can provide a fairly
  strong seed field ($\gsim 10^{-9} G$), then a kinematic dynamo can
   operate to amplify the fields to $\mu G$ strengths in a galactic
 disk. A sphere of radius $L$, encloses $(L/r_{sn})^3$ remnants where
  $r_{sn}$ is the size of a remnant. The rms value of the field at a
scale $L> r_{sn}$ would be roughly $\sim B_{sn} (r_{sn}/L)^{3/2}$. At
 the scale of the compact region, $L=r_0=100$pc, the large scale seed
		    field is of order $10^{-7}$ G.

\noindent{\em Angular momentum transport}

 There is an angular momentum transport mechanism required to form a
seed mass in a central region and one for the compact central mass to
       further collapse; the two schemes, in principle, can be
     different. Loeb \& Rasio (1994) concluded from hydrodynamic
  simulations that fragmentation occurs halting a direct collapse to
  relativistic scales and suggested that low spin systems could form
supermassive seeds of disk or star geometry which would contract under
    radiative viscosity.  The picture proposed here is that direct
    magnetic stress or gravitational instabilities in the disk can
 transport matter into a compact region on a dynamical timescale. The
 scheme of gravitational instabilities hinges on $Q_T<1$ in the disk
whereas the direct magnetic stress depends on the supernova efficiency
and field expulsion. The presence of high magnetic energy density from
 MHD turbulence has been invoked to explain high velocity dispersions
      of H I clouds in galactic disks (Sellwood \& Balbus 1999).

 Angular momentum in the inner region ($\lsim 1$ pc) is likely to be
    more complicated due to radiation pressure that would prevent
  collapse.  The model of self-gravitating magnetized collapse in \S
\ref{self} was illustrative of the timescale involved but a detailed
model for the inner region that takes into account radiation pressure,
radiative viscosity and evolution of the field is required.  If dynamo
  action takes place within the compact region ($\lsim$100 pc), the
    exponentiation timescale in the linear regime will be of order
  $1/\omega(r_0) \sim 10^6$ yrs. The residual large scale seed field
   ($\sim 10^{-7} G$) from supernovae can be amplified to dynamical
 values (mG) in $10^7$ years and the e-folding timescale for magnetic
braking is $r_0/v_A \sim (M(r_0)/r_0)^{1/2}/B =10^6$ yrs corresponding
		       to this field strength.
\section{Conclusions}
\label{conclude}

In this work, a semi-analytic model of quasar formation was attempted
which captures the essential details of star formation with supernovae
   feedback and angular momentum transport via magnetic fields and
self-gravity.  Based on the results in this work (as summarized in \S
\ref{disc}), and for the relevant range in parameter space, it is seen
   that a $10^{10} \Msun$ cloud at $z$ of 8 can collapse to form a
   gaseous disk in the dark halo by $z$ of 4.8 and about $10^7-10^8
\Msun$ of gas (which is 0.01--0.1 of the total disk mass) accretes
 into a $r_0 \lsim 100$ pc region in a fast timescale of $10^8$ yrs.
The magnetic stress from supernovae can be significant in the disk and
is preferred over other possibilities as observational evidence exists
for star formation, and large scale fields at $z \sim 2$. The collapse
 solutions do not hinge on the source of magnetic fields- SN or small
scale dynamo. The small scale dynamo needs a few rotation time scales
 to build up and that could be significant- $10^8$ yrs. The arguments
   for SN fields are given in section \S \ref{magnetic}; the strong
 fields suggested can cause fast accretion rates instantly after the
disk forms. The field strength estimation relies only on flux freezing
     arguments.  Further, the alternative collapse solution using
	       gravitational instabilty is also given.

  The solution to the non-linear diffusion equation presented in \S
\ref{bgd} can be applied in other contexts, eg. protostellar disks.
As a part of future work it is planned to investigate a detailed model
 of the inner region using a simulational approach, the impact of the
    restriction on the mass range of black hole hosts ($\cM-z$) by
   supernovae feedback on the details of the quasar luminosity function, dwarf galaxy formation, and the strength of the large scale seed field.
\begin{acknowledgements}
	 I thank K. Subramanian for discussions.  I thank the referee for helpful comments and suggestions.
\end{acknowledgements}
\appendix
\section{Core condensation in halos}
\label{core}
Here we calculate the change in gravitational potential energy from a
truncated isothermal halo to a truncated isothermal halo of stars and
dark matter containing an exponential gas disk.  The final halo mass,
 $ M_h= (1-f_g) \cM$ is taken to be due to stars and dark matter, and
  remaining mass is in the form of a gaseous disk with $M_d= \cM f_b
   (1-f_\ast)=f_g \cM$ where $f_g$ is the gas fraction. The initial
		     potential energy is given by
\begin{equation}
		    W_i= - {G ~\cM^2 \over 2 r_v}.
\end{equation}
		The final potential energy is given by
\begin{equation}
  W_f= - {G M_h^2 \over 2 r_v} -0.3 {G M_d^2 \over r_d} +{1 \over 2}
\int (\rho_d \Phi_h + \rho_h \Phi_d) \,\diff V,
\end{equation}
where the first two term represents the self potential energies due a
 truncated isothermal halo of size $r_v$, and an exponential disk of
size $r_d$, respectively, and the third term is due to the interaction
 between them. The first part of the last term with (the disk surface
 density $\Sigma(r) =\exp(-r/r_d)~ M_d/(2 \pi r_d^2) $ and $\Phi_h= G
	M_h/r_v$)can be reduced to $-(1/2) (G M_d M_h/r_v) (1+
\exp(-1/\lambda_v)/\lambda_v)$ where the term involving $\lambda_v$ is
	 negligible in the range of interest for which $0.1 >
\lambda_v > 0.05$. The second part of the interaction term can be evaluated 
  with the disk potential expressed as a Bessel series in the region
  $r<r_v$ where $\rho_h=M_h/(4 \pi r_v r^2)$ contributes. This term
      works out to be $-(G M_d M_h/r_v) 0.3/\lambda_v$ to a good
      approximation, where the collapse factor, $r_v/r_d \simeq
	     1/\lambda_v$, is in the range 10--20 (see \S
\ref{collapse}). Combining the two we obtain
\begin{displaymath}
      {1 \over 2} \int (\rho_d \Phi_h + \rho_h \Phi_d) \,\diff V
\approx -{G M_d M_h \over 2~ r_v} \left (1+{0.6 \over \lambda_v} \right). 
\end{displaymath}
   Finally, we obtain the change in gravitational potential energy
\begin{displaymath}
\Delta W(\cM, f_g, \lambda_v, r_v)= 
{G \cM^2 \over 2~r_v}
\end{displaymath}
\begin{displaymath}
 \times \left \{  (1-f_g)^2 
- 1+ f_g^2 \,{0.3 \over \lambda_v}+f_g (1-f_g)\left (1+{0.6 \over
 \lambda_v} \right ) \right \}
\end{displaymath} 
\begin{equation}
= {G \cM^2 \over 2~r_v}
 f_g \left [(2-f_g) {0.3 \over \lambda_v}-1 \right ]
\approx {G \cM^2 \over r_v} f_g \left ({0.3 \over \lambda_v}-{1 \over 2}
\right)
\end{equation}
	      which is bounded by $0.5 ~G \cM^2 / r_v$.
\section{Time-dependent evolution of an alpha disk}
\label{clg}
Here we first consider a simple polytrope and then discuss the case of
    a disk coupled with an local energy dissipation condition.  By
	assuming that the gas is a polytrope of index $\gamma$,
\begin{equation}
		       P=K~(\Sigma/2H)^\gamma,
\end{equation}
 and using the vertical equilibrium, eqn (\ref{vertical}), we obtain
\begin{eqnarray}
\Pi^{ss}_{r\phi}&=& -2 \alpha_{ss} P H \nonumber \\&=&
-\alpha_{ss} K^{2/(\gamma+1)} \Sigma^{(3 \gamma+1)/(\gamma+1)}
\omega^{2(\gamma-1)/(\gamma+1)}.
\end{eqnarray}
  In some applications it is appropriate to consider a disk with an
  energy dissipation condition as has been done by Cannizzo, Lee \&
   Goodman 1990 (CLG), for the case of disk accretion of a tidally
    distrupted star onto a massive black hole. Using the solution
presented in \S \ref{bgd}, we can generalize to a case of a rotational
 law, $\omega \propto r^a$, and viscous stress, $\Pi_{r\phi} \propto
\Sigma^b r^c$. The viscous stress is given by
\begin{equation}
\Pi_{r\phi}= -2 \alpha_{ss} P H= -{\alpha_{ss} k_B T \over \mu m_p} \Sigma.
\label{avis}
\end{equation}
    where $T$ refers to the central temperature and $P$ is the gas
pressure.  Further, in the general prescription, the radiative flux is
		matched to the viscous dissipation by
\begin{equation}
\alpha_{ss} \omega H P |a|= {4 \over 3} ~{4 \sigma T^4 \over \kappa \rho H}
\end{equation}
 where $\sigma$ is the Stefan-Boltzmann constant and $\kappa(\rho,T)$
		   is the opacity. This reduces to
\begin{equation}
  T^3= {3 \over 16} \left ( {\alpha_{ss} k_b \over \mu m_p} \right )
\omega \Sigma^2 {\kappa \over 4 \sigma} |a|
\end{equation}
	Now, putting this back in eqn (\ref{avis}), we obtain
\begin{eqnarray}
\Pi_{r\phi}&=& -\left ({3 \over 16}\right)^{1/3} \left ( {\alpha_{ss} k_b \over \mu m_p} \right )^{4/3}\nonumber \\
&& \times \omega^{1/3} \Sigma^{5/3} \left (\kappa \over 4 \sigma \right)^{1/3} 
|a|^{1/3}
\end{eqnarray}
This implies $b=5/3, c=a/3$ for a constant $\kappa$. For the case of a
 Kepler potential we obtain the scaling (and opacity due to electron
scattering which is a constant) used in CLG ($a=-3/2, c=-1/2, b=5/3$)
    and their solution is given by eqn (\ref{solution}) for these
 indices. The case of a flat rotation law and opacity due to electron
scattering yields ($a=-1, b=5/3, c=-1/3$). This solution can be easily
extended to more general opacities of the form $\kappa \propto \rho^p
   T^q$, such as the Kramer's, and the appropriate solutions can be
			   easily found.

\end{document}